\documentclass[referee, sn-nature]{sn-jnl}

\usepackage{graphicx}%
\usepackage{multirow}%
\usepackage{amsmath,amssymb,amsfonts}%
\usepackage{amsthm}%
\usepackage{mathrsfs}%
\usepackage[title]{appendix}%
\usepackage{xcolor}%
\usepackage{textcomp}%
\usepackage{manyfoot}%
\usepackage{booktabs}%
\usepackage{algorithm}%
\usepackage{algorithmicx}%
\usepackage{algpseudocode}%
\usepackage{listings}%

\usepackage{physics}

\newcommand{\ms}[1]{\mbox{\scriptsize #1}}
\newcommand{\msi}[1]{\mbox{\scriptsize \textit{#1}}}

\begin{document} 

\title[Revealing spoofing of classical radar using quantum noise]{Revealing spoofing of classical radar using quantum noise}
\author*[1]{\fnm{Jonathan N.} \sur{Blakely}}\email{jonathan.n.blakely.civ@army.mil}

\author[1]{\fnm{Shawn D.} \sur{Pethel}}

\author[2]{\fnm{Kurt} \sur{Jacobs}}

\affil*[1]{\orgname{U. S. Army DEVCOM Aviation \& Missile Center}, \orgaddress{ \city{Redstone Arsenal}, \postcode{35898}, \state{Alabama}, \country{USA}}}

\affil[2]{ \orgname{U. S. Army DEVCOM Army Research Laboratory}, \orgaddress{ \city{Adelphi}, \postcode{20783}, \state{Maryland}, \country{USA}}}

\abstract{
Electromagnetic remote sensing technologies such as radar can be misled by targets that generate spoof pulses. Typically, a would-be spoofer must make measurements to characterize a received pulse in order to design a convincing spoof pulse. The precision of such measurements is ultimately limited by quantum noise. Here we introduce a model of electromagnetic spoofing that includes effects of practical importance that were neglected in prior theoretical studies. In particular, the model includes thermal background noise and digital quantization noise, as well as loss in transmission, propagation, and reception. We derive the optimal probability of detecting a spoofer allowed by quantum physics. We show that heterodyne reception and thresholding closely approaches this optimal performance. Finally, we show that a high degree of certainty in spoof detection can be reached by Bayesian inference from a sequence of received pulses. Together these results suggest that a practically realizable receiver could plausibly detect a radar spoofer by observing errors in the spoof pulses due to quantum noise.}


\maketitle

\section{Introduction}
It was recently shown that quantum mechanics fundamentally limits the ability to spoof electromagnetic pulses to fool a sensor~\cite{blakely2022quantum}. Specifically, the measurement made by an adversary to characterize a pulse is generally insufficient to fully determine its quantum state. Thus, in principle, a friendly receiver can use knowledge of the transmitted quantum state to detect spoofs. A classic application of spoofing is where an airborne target emits spoof pulses to avoid being tracked by a ground-based radar~\cite{schleher1999electronic, genova2018electronic}. Spoofing also has non-adversarial applications in hardware-in-the-loop testing~\cite{strydom2012hardware, strydom2014high, heagney2018digital}. A limitation of the work in ref.~\cite{blakely2022quantum} was the neglect of important practical considerations such as noise and loss. Clearly, a full understanding of the importance of quantum physics to real world spoofing requires a model that includes these effects. Here we introduce such a model including both thermal background noise and digital quantization noise, as well as loss in transmission, propagation, and reception.

The model provides insight into the relative importance of these effects in comparison to the purely quantum limits on spoofing previously identified. We analyze the performance of a quantum optimal receiver in discriminating spoofs. We find that, on one hand, loss and thermal noise degrade the ability to detect spoofing, while on the other hand, quantization noise in the spoof pulses acts similarly to quantum noise thus increasing the ability to discriminate. Finally, we examine a realizable receiver architecture, heterodyne reception combined with a thresholding procedure, which is shown to closely approach quantum optimal performance. Altogether, these results suggest that even under realistic conditions of large loss and background noise a realizable receiver can detect spoofing errors due to quantum noise. To be clear, quantum noise-based spoof detection is not a practical approach to current spoofing technologies. These devices introduce a variety of errors and a quantity of classical noise that provide the basis for existing spoof detection methods \cite{schleher1999electronic, genova2018electronic}. Rather, this work is forward looking to a future spoofing technology that can mimic a transmitted pulse with an accuracy approaching the quantum limit \cite{blakely2022quantum}.

\begin{figure}[bht]
\includegraphics{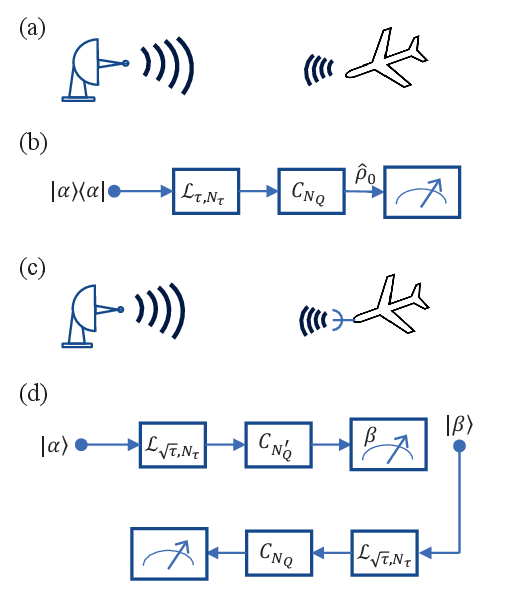}
\caption{Schematic depictions of the hypotheses to be discriminated where (a) and (c) illustrate a true echo from a target and a target-generated spoof, respectively, and (b) and (d) show the quantum channel models of each hypothesis. See text for more explanation }
\label{fig:channels}
\end{figure}

We introduce our model of spoof detection in Sec.~\ref{model}. The model takes the form of a quantum hypothesis test deciding between the  presence or absence of a spoofer. In Sec.~\ref{Optimal_Detection}, we determine the quantum optimal probability of discriminating between the hypotheses and present a specific architecture for realizing optimal detection. In Sec.~\ref{Heterodyne_Reception}, we analyze a more practically realizable detection scheme based on heterodyne reception and thresholding. In Sec.~\ref{Example}, we examine a specific radar application where detection using heterodyne detection closely approaches optimal performance. In Sec.~\ref{Bayes}, we show how Bayesian inference can be used to aggregate information from multiple received pulses to detect spoofing with near certainty. Lastly, in Sec.~\ref{Conclusion}, we give concluding remarks.

\section{Results}
\subsection{A quantum model of radar spoofing}
\label{model}

We model spoofing as a hypothesis test undertaken by the operator of a friendly receiver who must decide if a received pulse in a specific range-Doppler bin is a true reflection from a target of interest (hypothesis $H_0$), or a spoof pulse generated by an adversary (hypothesis $H_1$). We assume the target is probed by a narrowband, transform-limited pulse represented by a coherent state of a single, generalized, temporal mode (ignoring consideration of the spatial field pattern, for simplicity). By design, the amplitude $\alpha$ of the transmitted coherent state is a complex random variable with the zero-mean Gaussian probability density given by 
\begin{align}
P(\alpha) = \frac{\lambda}{\pi}e^{-\lambda |\alpha|^2},
\label{alpha_pdf}
\end{align}
where $\lambda$ is a positive constant. The value of $\alpha$ is assumed to be known by the operator, but not by the adversary.

Under hypothesis $H_0$, the received pulse is a true reflection off a target of interest, as depicted in Fig. \ref{fig:channels}(a). In this case, the pulse suffers loss as it is radiated from a transmitting source (e.g. an antenna or a laser) with some degree of impedance mismatch, propagated out to the target and back, and is received by a detector (e.g. an antenna or photodetector). Thermal noise is added to the signal at transmission, reflection, and reception. We model these processes by a single-mode, lossy, Gaussian bosonic channel $\mathcal{L}_{\tau, N_T}$ with total transmissivity $\tau$ and mean noise photon number $N_T$ \cite{weedbrook2012gaussian}. The action of $\mathcal{L}_{\tau, N_T}$ on an input Gaussian state with displacement vector $\mathbf{x}$ and covariance matrix $\mathbf{V}$ is the transformation
\begin{align}
\mathbf{x} &\rightarrow \sqrt{\tau}\mathbf{x}, \\   \mathbf{V} &\rightarrow \tau \mathbf{V} + (1-\tau)\left(2N_T+1\right)\mathbf{I}. 
\end{align}
In the transformation of the covariance matrix, the first term represents the reduction of the size of fluctuations due to loss processes, while the second term represents fluctuations added by thermal noise. In what follows, it will be useful to let $N_T=N'_T/(1-\tau)$ where $N'_T$ is a fixed mean noise photon number independent of $\tau$. 

Quantization noise is added upon digitization of the received signal. Typically, quantization noise in high resolution digitization is modeled as uniformly distributed over the range $E$ corresponding to the least significant bit, with zero mean and variance $E/12$ \cite{bennett1948spectra}. For analytical convenience, it is here assumed that the quantization process is a classical Gaussian noise channel $\mathcal{C}_{\xi}$ that adds Gaussian noise with variance $\xi = E/12$ to the input signal. The action of $\mathcal{C}_{\xi}$ on an input Gaussian state with displacement vector $\mathbf{x}$ and covariance matrix $\mathbf{V}$ is the transformation $\mathbf{x} \rightarrow \mathbf{x}$,  $\mathbf{V} \rightarrow \mathbf{V} + \xi\mathbf{I}$  \cite{weedbrook2012gaussian, zamir1996lattice}. 

The complete model under hypothesis $H_0$, including the final measurement made by the receiver, is depicted in Fig. \ref{fig:channels}(b). Assuming the transmitted state is $\hat{\rho} = \ket{\alpha}\bra{\alpha}$, for which 
\begin{align}
& \mathbf{x}=  \left [ \begin{matrix}
& \alpha + \alpha^*   \\
& i \left( \alpha^* - \alpha\right)   
\end{matrix}
\right  ],
\label{displacement_vector}
\end{align}
and $\mathbf{V} =\mathbf{I}$, where $\mathbf{I}$ is the identity matrix, the state measured by the receiver under hypothesis $H_0$, i.e. $\mathcal{C}_{\xi}\left(\mathcal{L}_{\tau, N_T}\left(\hat{\rho}\right)\right)$, has displacement vector 
\begin{align}
& \mathbf{x}_0= \sqrt{\tau} \mathbf{x}  
\label{displacement_vector_0}
\end{align}
and covariance matrix
\begin{align}
\mathbf{V}_0 = \left\{2N_0+1\right \}\mathbf{I},
\label{covariance_matrix_0}
\end{align}
where 
\begin{align}
    N_0 = N'_T+\xi/2 
\end{align}
and $\mathbf{I}$ is the identity matrix. The `$0$' subscripts in Eqs.(\ref{displacement_vector_0}) and (\ref{covariance_matrix_0}) indicate that these quantities describe the quantum state at the receiver under hypothesis $H_0$. Equivalently, this state can be represented by the density operator
\begin{align}
\hat{\rho}_0 = \frac{1}{\pi N_0}\int d^2 \alpha' e^{-\frac{|\alpha' - \sqrt{\tau}\alpha|^2}{N_0}} \ket{\alpha'}\bra{\alpha'} .
\label{density_H0}
\end{align}

Under hypothesis $H_1$, the received pulse is a spoof, as represented in Fig.~\ref{fig:channels}(c). We assume the spoof is generated by an adversary who has performed a single measurement on the transmitted state and aims to reproduce this state as closely as possible. We refer to this measure-and-prepare approach as \em classical \em spoofing~\cite{blakely2022quantum}. We model propagation from the transmitter to the spoofer, and from the spoofer to the receiver as two separate passes through the lossy channel $\mathcal{L}_{\sqrt{\tau}, N_T}$, which effects the transformation $\mathbf{x} \rightarrow \tau^{1/4} \mathbf{x}$,   $\mathbf{V} \rightarrow \sqrt{\tau} \mathbf{V} + (1-\sqrt{\tau})\left(2N_T+1\right)\mathbf{I}$. When the output of the first channel is fed directly to the second channel, the result is equivalent to the single channel under hypothesis $H_0$, i.e. $\mathcal{L}_{\sqrt{\tau}, N_T}\left(\mathcal{L}_{\sqrt{\tau}, N_T}\left(\hat{\rho}\right)\right)=\mathcal{L}_{\tau, N_T}\left(\hat{\rho}\right)$. Thus, if the adversary were able to exactly copy the transmitted quantum state, the receiver would have no basis for discriminating a spoof from a real return. However, quantum physics does not allow the adversary to fully characterize the transmitted state with a single measurement. 

The optimal single measurement for estimation of the Gaussian-distributed mean amplitude of a noisy coherent state such as is received by the adversary, is heterodyne detection~\cite{guctua2010quantum}. Thus, we assume the adversary makes a heterodyne measurement of the complex amplitude. Heterodyne detection has a long history in quantum optics, but is also essentially the operation performed by a coherent radar receiver insofar as the received signal is mixed down to an intermediate frequency and then input to a quadrature detector and matched filters that output the real and imaginary parts of the complex amplitude. We further allow for the introduction of quantization noise by the adversary as the quadrature signals are typically digitized. 

Ideal heterodyne detection realizes the positive operator-valued measure with measurement operators $\ket{\beta}\bra{\beta}/\sqrt{\pi}$ \cite{shapiro1984phase}. The statistics for heterodyne measurement on the output of the lossy channel representing propagation from the transmitter to the spoofer with added quantization noise, i.e. $\mathcal{C}_{\xi'} \left(\mathcal{L}_{\sqrt{\tau},N_T}\left(\hat{\rho}\right)\right)$, are described by the probability density
\begin{align}
P(\beta) &= \trace \left [ \frac{\ket{\beta}\bra{\beta}}{\pi} \mathcal{C}_{\xi'} \left(\mathcal{L}_{\sqrt{\tau},N_T}\left(\hat{\rho}\right)\right) \right] \\
&= \frac{\exp \left(-\frac{|\tau^{1/4} \alpha - \beta|^2}{\left(1+\sqrt{\tau} \right)^{-1} N'_T + \xi'/2 + 1}\right)}{\pi \left[ \left(1+\sqrt{\tau} \right)^{-1}N'_T + \xi'/2 + 1\right ]} 
\label{adversary_stats}
\end{align} 
where $\beta$ is the complex measurement outcome. The variance of the additive quantization noise is $\xi'$, which is generally not equal to that of the friendly receiver, $\xi$. The quantization noise levels are different for these two receivers because they are typically receiving signals of very different amplitudes.

The adversary generates a spoof pulse in the same generalized temporal mode with complex amplitude $\beta$ and it passes through the lossy channel $\mathcal{L}_{\sqrt{\tau},N_T}$ representing the path from the adversary to the friendly receiver. The receiver is assumed to introduce quantization noise upon reception, resulting in the state $\mathcal{C}_{\xi}\left(\mathcal{L}_{\sqrt{\tau},N_T}\left(\ket{\beta}\bra{\beta}\right)\right)$. It is assumed that the receiver knows the adversary's measurement statistics, but not the measurement outcome $\beta$. Thus, the state of the pulse at the receiver is a mixture of coherent states weighted by the density Eq.(\ref{adversary_stats}) as expressed by the displacement vector 
\begin{align}
& \mathbf{x}_1= \mathbf{x}_0,
\label{displacement_vector_1}
\end{align}
and the covariance matrix
\begin{align}
\mathbf{V}_1 = \mathbf{V}_0 + 2\sqrt{\tau}\left(1+\xi'/2\right)\mathbf{I}.
\label{covariance_matrix_1}
\end{align}
The subscripts in Eqs.(\ref{displacement_vector_1}) and (\ref{covariance_matrix_1}) indicate that these quantities describe the quantum state under hypothesis $H_1$. Equivalently, this state can be represented by the density operator
\begin{align}
\hat{\rho}_1 = \frac{1}{\pi N_1}\int d^2 \alpha' e^{-\frac{|\alpha' - \sqrt{\tau}\alpha|^2}{N_1}}  \ket{\alpha'}\bra{\alpha'} 
\label{density_H1}
\end{align}
where 
\begin{align}
    N_1 = N'_T + \xi/2 +\sqrt{\tau}\left(1+\xi'/2\right) .
\end{align}

Upon reception, a decision must be made as to whether a received pulse is most consistent with the state specified by Eqs.(\ref{displacement_vector_0}) and (\ref{covariance_matrix_0}) under hypothesis $H_0$ or by Eqs.(\ref{displacement_vector_1}) and (\ref{covariance_matrix_1}) under hypothesis $H_1$. Comparing Eqs.(\ref{displacement_vector_0}) and (\ref{displacement_vector_1}), it can be concluded that the displacement vector provides no basis for a decision because it is the same under both hypotheses. The second term on the right hand side of Eq.(\ref{covariance_matrix_1}) does provide a basis for a decision. The first term in parentheses in this equation represents the quantum noise in the heterodyne measurement outcome. One half of this noise is attributable to quantum noise in the transmitted coherent state. The other half is quantum noise associated with the Heisenberg uncertainty relation between the real and imaginary field quadratures in the course of an ideal heterodyne measurement. The second term in parentheses in Eq.(\ref{covariance_matrix_1}) represents the noise added by the adversary through digital quantization. Interestingly, the adversary's quantum and classical noise enter the discrimination problem in the same manner even though their physical origins are distinct.

Having now framed spoof detection as a hypothesis test, we next turn to the analysis of specific measurement strategies that the receiver operator might adopt when seeking to detect the presence of a spoofer. In the sections that follow, the optimal measurement strategy allowed by quantum mechanics will be examined, as well as a practically realizable strategy that closely approaches the optimum. 

\subsection{Quantum Optimal Detection of Spoofing}
\label{Optimal_Detection}
Quantum detection theory enables the calculation of the probability of successful detection assuming the receiver executes the measurement and decision criterion that minimizes the Bayesian total probability of error over all positive operator-valued measures \cite{1976quantum}. In this section, we examine this optimal performance and the receiver architecture that would achieve it. Throughout this section, the Bayesian prior probability that a pulse is a spoof is assumed to be 0.5. It is straightforward to generalize the results that follow to allow for other values of this probability, but for the sake of clarity, only the one case will be discussed. Letting $P_{\ms{opt}}$ denote the probability of choosing the hypothesis that corresponds to the truth using the optimal receiver, then
\begin{align}
P_{\ms{opt}} = \frac{1}{2} \left(1+ \frac{1}{2}||\hat{\rho}_1-\hat{\rho}_0||_1\right),
\label{P_success}
\end{align}
assuming equal Bayesian prior probabilities for the two hypotheses, equal costs for all types of error, and where $|| \cdot ||_1$ denotes the trace norm \cite{1976quantum}. 

We can obtain a fairly simple expression for $P_{\ms{opt}}$ by noting that it is unchanged if we apply a unitary transformation to both $\hat{\rho}_0$ and $\hat{\rho}_1$. Since according to Eq.(\ref{displacement_vector_1}) both states have the same displacement vector (phase space centroid), we can apply a displacement transformation to reduce the displacement vectors of both to zero while leaving the variances unchanged. This unitary transformation does not affect $P_{\ms{opt}}$, but the resulting states are then thermal states and are thus diagonal in the Fock basis. 
%
Following Helstrom \cite{helstrom1967detection}, the optimal probability of successful discrimination for any value of $\alpha$ is then 
\begin{align}
P_{\ms{opt}} & = \frac{1}{2} \frac{1}{N_0+1} \sum_{n=0}^m \left(\frac{N_0}{N_0+1}\right)^n \nonumber \\
& + \frac{1}{2}   \frac{1}{N_1+1} \sum_{n=m}^\infty \left(\frac{N_1}{N_1+1}\right)^n ,
\label{P_opt_expression} 
\end{align} 
with
\begin{align}
m & =  \mbox{floor} \left\{ \frac{\ln \frac{N_1+1}{N_0+1} }{ \ln \left[\frac{N_1(N_0+1)}{N_0(N_1+1)}\right] } \right\}.
\label{optimal_prob}
\end{align}  

For the $\alpha = 0$ case, Helstrom found optimal discrimination could be performed by photon counting followed by comparison to a threshold of value $m$ \cite{helstrom1967detection}. It follows that for $\alpha \neq 0$, optimal discrimination can be performed by a receiver that first displaces the received signal by $\alpha$ and then counts photons and compares to the threshold. In the context of microwaves, the displacement can be realized by homodyne down conversion. In principle, photon counting could be done on the resulting baseband signal. Unfortunately, existing single photon detectors in the microwave regime have low quantum efficiencies \cite{chen2011microwave, pankratov2022towards}. Thus, we next analyze heterodyne detection and thresholding, a currently realizable architecture. Importantly, this approach will be shown to perform close to optimally.

\subsection{Detection of Spoofing with Heterodyne Reception}
\label{Heterodyne_Reception}
Consider a receiver that makes a heterodyne measurement whose outcome is a complex amplitude that is compared to a threshold to discriminate the two hypotheses. Under hypothesis $H_k$, with $k=0,1$, the heterodyne measurement outcome $\beta$ is a random variable with probability density~\cite{shapiro1984phase}
\begin{align}
P(\beta |H_k) &= \trace \left( \frac{\ket{\beta}\bra{\beta}}{\pi} \hat{\rho}_k \right) =  \frac{  e^{-\left| \beta - \sqrt{\tau}\alpha \right|^2/(N_k + 1)}}{\pi\left( N_k + 1\right)} ,
\label{het_stats_0}
\end{align}
We introduce a threshold $\mu$ such that if $|\beta| \le \mu$ we select hypothesis $H_0$, and conversely if $|\beta | > \mu$ we  select hypothesis $H_1$. The set of $\beta$ values satisfying the former condition, which we will refer to as $Z_0$, is a filled circle (a disk) with radius $\mu$ centered on $\sqrt{\tau}\alpha$. The set satisfying the latter condition, referred to as $Z_1$, is the rest of the complex plane. The probability of success in choosing the true hypothesis, $P_\text{het}$, is the sum of the probability of choosing $H_0$ when it is true and the probability of choosing $H_1$ when it is true. Mathematically, this is 
\begin{align}
P_\text{het} & = \frac{1}{2}\int  \limits_{Z_0} d^2 \beta P(\beta |H_0) + \frac{1}{2}\int  \limits_{Z_1} d^2 \beta P(\beta |H_1) \\
& = \frac{1}{2}\left ( 1- e^{-\mu^2/(N_0+1)}\right) + \frac{1}{2} e^{-\mu^2/(N_1+1)},
\end{align}
where, again, an assumption of equal prior probabilities has been made. It follows that the value of the threshold $\mu$ that optimizes $P_\text{het}$ is equal to the magnitude of $\beta$ where the curves $P(\beta |H_0)$ and $P(\beta |H_1)$ intersect. Specifically, the optimal threshold is
\begin{align}
\mu_\text{opt} =  \sqrt{\frac{N_0 + 1}{1-\frac{N_0 + 1}{N_1 + 1}}  \ln \left(\frac{N_1 + 1}{N_0 + 1}\right)} .
\end{align}
In the next section, we will compare this detection scheme with optimal detection in a specific application.

\subsection{An Example}
\label{Example}

As a specific example, we use the parameters of a W-band radar defined in Refs. \cite{zhuang2022ultimate, wu2022entanglement} where
\begin{align}
\tau = \left( \frac{G_T}{4\pi R^2}\right)\left(\frac{\sigma A_R}{4\pi R^2}\right), 
\end{align}
with $G_T= A_R/\left(2 \pi c/\omega_0 \right)^2$. Here $G_T$ is the radar antenna gain, $A_R=1$ $\text{m}^2$ is its effective area, $\sigma = 0.01$ $\text{m}^2$ is the target cross section, $\omega_0/2\pi = 100$ GHz is the pulse center frequency, and $c$ is the speed of light. The mean noise photon number $N'_T = 32$, corresponding to a receiver noise temperature of $150$ K. 

To determine the magnitude of the quantization noise, we note that according to Eq. (\ref{alpha_pdf}) the real and imaginary parts of the mean complex amplitude of the transmitted pulse are  zero mean random variables with variance $(2\lambda)^{-1}$. So the average mean photon number in such pulses is $(2\lambda)^{-1}$. The signal under hypothesis $H_0$ passes through the channel $\mathcal{L}_{\tau, N_T}$ before arriving at the receiver. The signal would emerge from this channel with an average mean photon number $\tau/(2\lambda)$. We assume this signal is quantized at the receiver with $n$ bits of resolution such that the least significant bit corresponds to a range $E \approx 2^{-n}\tau/(2\lambda)$ with units of photon number. The variance of the quantization noise is then taken to be $\xi = E/12$. The value of $(2\lambda)^{-1}$ is chosen by assuming the pulse width $T = 1 \mu$s,  and the average power $P_\text{ave} = 10$ kW, giving an average pulse energy of $10^{-2}$ J. Under the assumption of narrow bandwidth, the energy per photon is approximately $\hbar \omega_0$. Then the effective mean photon number for quantization noise at the radar receiver is
\begin{align}
\xi \approx \tau\frac{2^{-n} T  P_\text{ave}}{12 \hbar \omega_0 }
\end{align}
A common value for $n$ in existing microwave technology is 10, giving $\xi \approx 9 \times 10^{4}$ at a range of 1 km. But due to \begin{figure}[tbh]
\includegraphics{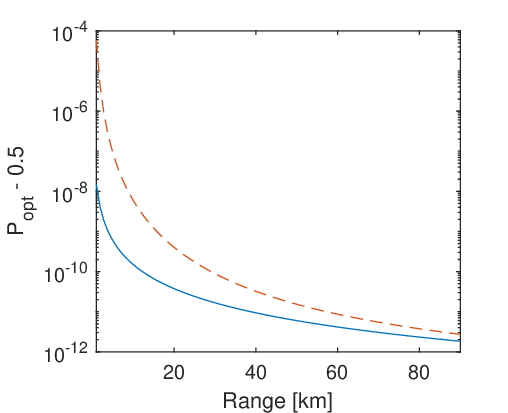}
\caption{The probability of successfully discriminating between true and spoofed pulses for an optimal receiver with quantization noise due to digitization with bit resolution $n=32$  (dashed red line) and with no quantization noise (solid blue line). }
\label{fig:P_opt}
\end{figure}the rapid increase of loss with increasing range, $\xi$ drops to approximately 1 at 17 km. Beyond this level of resolution, the quantization noise is small compared to the quantum noise in this model. Existing ultra high resolution analog-to-digital converters can have $n=32$, giving $\xi \approx 1$ at just 375 m.  

The quantization noise introduced by the spoofer will necessarily have larger variance than $\xi$ because the spoofer receives the signal after passing through the less lossy channel $\mathcal{L}_{\sqrt{\tau}, N_T}$. By the same reasoning as above
\begin{align}
\xi' \approx \sqrt{\tau}\frac{ 2^{-n} T P_\text{ave}}{12 \hbar \omega_0 }.
\end{align}
In this case, with $n=10$, $\xi'$ falls to approximately 1 at the impractical distance of 180,000 km, and with $n=32$, $\xi' \approx 1$ at 88 km. Beyond this range, the spoofer can be said to be limited chiefly by quantum noise.

With all the model parameters now set, we first examine the performance of optimal spoof detection. The optimal probability of successful discrimination, as given by Eq.(\ref{optimal_prob}), is shown as a function of range in Figure \ref{fig:P_opt}. Since the prior probability of spoofing is 0.5, the probability of successful discrimination before transmitting any signal is also 0.5. Thus, in the figure 0.5 is subtracted from $P_\text{opt}$ to emphasize the increase due to the gain of information from reception and measurement of a pulse. The blue line is the probability with infinite bit resolution, i.e., $\xi = \xi'=0$.  The non-zero value (after subtracting 0.5) indicates that, in principle, quantum noise alone provides a sufficient physical basis for detecting the spoofer. Importantly, since the spoofer is assumed to employ the quantum optimal measurement for estimating the transmitted quantum state, no other measure-and-prepare \begin{figure}[tbh]
\includegraphics{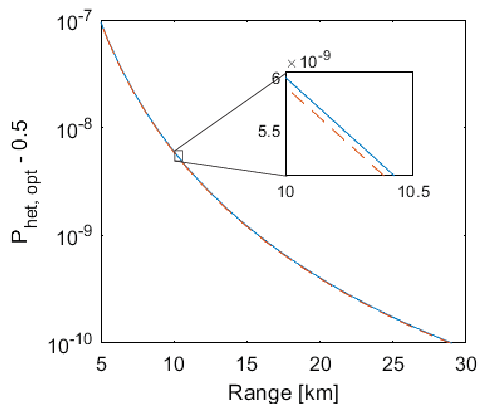}
\caption{The probability of successfully discriminating between true and spoofed pulses with quantization noise due to digitization with bit resolution $n=32$ for an optimal receiver (solid blue line) and a heterodyne receiver with threshold detection (dashed red line). }
\label{fig:Hel_het}
\end{figure}strategy can be devised to eliminate this physical basis. Thus, exploitation of quantum noise here provides a kind of quantum advantage in spoof detection.

The dashed, red line in Fig. \ref{fig:P_opt} is the success probability with a bit resolution $n = 32$ bits. As range increases, the $n=32$ probability approaches the $n=0$ probability. This trend illustrates the decreasing relative importance of classical quantization noise versus quantum noise at long ranges. 

Optimal performance can be compared to that of heterodyne reception and thresholding. For $n=32$, the probability, $P_\text{opt}$, is shown (solid blue line) along with the corresponding success probability for heterodyne reception, $P_\text{het}$, (dashed red line) in Fig. \ref{fig:Hel_het}. Importantly, the more practical heterodyne detection scheme closely approaches the performance of optimal detection. The inset shows how the former falls just short of the latter. 

With either detection method, the success probability is very small at most ranges. For example, at a range of 10 km, $P_\text{het}-0.5$ for this receiver is approximately $10^{-8}$. One might conclude that the increase in success probability over the prior probability would be too small to be of practical use in many applications. However, even a very small increase can be exploited by aggregating information from multiple transmissions through a process such as Bayesian inference, as described in the following section \cite{blakely2022quantum}.

\subsection{Bayesian Inference from Multiple Pulses}
\label{Bayes}
 The small effect of quantum noise added by an adversary can be exploited by aggregating the information collected from multiple pulses, each with a different random amplitude. Previously, Bayesian inference was used\begin{figure}[tbh]
\includegraphics{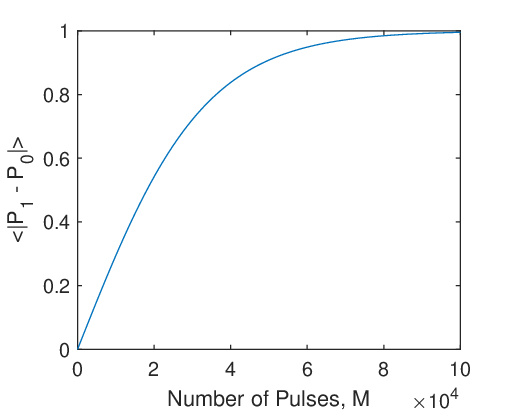}
\caption{Mean difference in prior probabilities as a function of number of pulses received. The prior probabilities are updated after each new pulse is received according to the procedure of Bayesian inference. The range is 1 km and the bit resolution is 32.}
\label{fig:Bayesian_conv}
\end{figure} to update the prior probabilities in a noise-free, loss-free, spoofing model for a binary phase shift keying signal set \cite{blakely2022quantum}. Here we apply the same approach to the current model of spoofing with heterodyne reception and threshold detection. 
 
Bayesian inference involves updating the prior probability after each new measurement outcome  \cite{winkler2003introduction}. Let $P_0$ ($P_1$) be the prior probability of hypothesis $H_0$ ($H_1$), respectively, after $M$ measurements. It is shown in Sec. \ref{Methods} that the difference between the prior probabilities after $M \gg 1$ trials will on average take the value
\begin{align}
\langle |P_1 - P_0 |\rangle \approx \frac{|1-e^{M\Delta_0 (\Delta_0 - \Delta_1)}|}{1+e^{M\Delta_0 (\Delta_0 - \Delta_1)}},
\label{difference_of_priors}
\end{align}
where
\begin{align}
\Delta_0 = 2e^{-\mu^2/(N_0 + 1)}-1,
\label{delta_-}
\end{align}
and
\begin{align}
\Delta_1 = 2 e^{-\mu^2/(N_1 + 1)} - 1.
\label{delta_+}
\end{align}
This approximation to the mean difference in probabilities as a function of $M$ is shown to approach unity at large $M$ in Fig. \ref{fig:Bayesian_conv} for the example parameters of Sec. \ref{Example} (and, in particular, $n = 32$). This result means that certainty is approached by one of the two hypotheses when enough pulses have been received. For example, $\langle |P_1 - P_0 |\rangle > 0.95$ after about $6 \times 10^5$ pulses. To achieve a desired value of $\langle |P_1 - P_0 |\rangle$ near one, the required number of samples is
\begin{align}
    M \approx \frac{1}{\Delta_0 (\Delta_0 - \Delta_1)} \ln \frac{1+\langle |P_1 - P_0 |\rangle }{1-\langle |P_1 - P_0 |\rangle}.
\end{align}
\begin{figure}[tbh]
\includegraphics{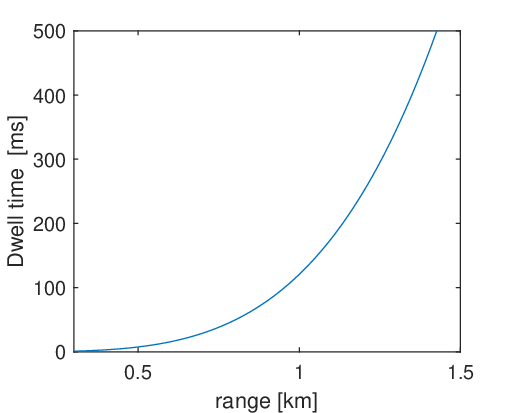}
\caption{Dwell time to reach a mean difference in prior probabilities of 0.9 as a function of range with a pulse repetition rate of 500 kHz.}
\label{fig:dwell_time}
\end{figure}Dividing this number by a pulse repetition rate would give the required dwell time on target to achieve a desired average level of certainty. Figure \ref{fig:dwell_time} shows the required dwell time as a function of range for the example parameters assuming a desired $\langle |P_1 - P_0 |\rangle$ of 0.9 and a pulse repetition rate of 500 kHz. At 1 km, the required dwell time is about 120 ms. During such an interval, a target with a velocity as high as $10^2$ m/s would not move by significant fraction of the range of 1 km.

\section{Discussion}
\label{Conclusion}
In this article, we have shown that a practically realizable receiver could plausibly detect a radar spoofer by observing errors in the spoof pulses due to quantum noise. In practice, information from many pulses would have to be aggregated to reach a meaningful degree of certainty, but in an example application this requirement was shown to be achievable. This exploitation of quantum noise constitutes a kind of quantum advantage in spoof detection.

To arrive at these results, we introduced a new model of radar spoofing that includes noise and loss. Key assumptions of the model were (1) the set of signals used by the radar (specifically, coherent states with Gaussian-distributed amplitudes), and (2) the limitation of the spoofer to a measure-and-prepare strategy. Extensions of this work could explore the consequences of modifying either of these assumptions. On the one hand, expanding the set of possible signals which the spoofer must discriminate could enhance the radar operator's ability to detect the spoofer. On the other hand, spoofing strategies that exploit more of the information available in the received quantum state than is extracted by a single measurement might allow for more deceptive spoofing. Our current work is pursuing both of these threads.

\section{Methods}
\label{Methods}
\subsection{Convergence of Bayesian Inference}
Here we derive Eq.(\ref{difference_of_priors}) assuming the radar transmits $M$ pulses, each with an independent, randomly chosen amplitude. Under either hypothesis, the radar operator's measurement has two possible outcomes, a determination that the received pulse is either a true return or a spoof. Let the symbols $-$ and $+$ indicate the measurement outcomes corresponding to a true return and a spoof, respectively. In general, if $H_i$ is true (where $i$ is either 0 or 1), then the probabilities of the two outcomes are 
\begin{align}
     P(\pm|H_i) = \frac{1}{2} \pm \frac{\Delta_i}{2} = \frac{1}{2} \left( 1 \pm \Delta_i \right), 
\end{align}
where $-1 \leq \Delta_i \leq 1$, and the specific value of $\Delta_i$ depends on the particular choice of measurement. These probabilities are known as \em likelihood \em functions. For heterodyne detection and thresholding, 
\begin{align}
     P(\pm|H_i) = \int  \limits_{Z_i}  P(\beta |H_i) \; d^2 \beta, 
\end{align}
from which follows
$\Delta_0$ and $\Delta_1$ as given by Eqs.(\ref{delta_-}) and (\ref{delta_+}), respectively.

Let the prior probability of the hypothesis $H_i$ before the first pulse is received be $P_0(H_i)$. After the $n^\text{th}$ pulse is received and measured, our new state-of-knowledge is obtained by multiplying the prior probabilities by the corresponding likelihood function and normalizing the result \cite{winkler2003introduction}. If we leave off the normalization (which we can always do after all $M$ measurements have been made) the prior probabilities after the $n^\text{th}$ pulse is received are 
\begin{align}
       P_n(H_i) &= 
       P(\pm|H_i) P_{n-1}(H_i),  \\
&= \frac{1}{2} \left( 1 \pm \Delta_i  \right) P_{n-1}(H_i), \\
& \approx  \frac{1}{2} \exp \left( \pm \Delta_i   \right) P_{n-1}(H_i), 
\end{align}
where the approximation in the last line is valid to the extent that $\Delta_i \ll 1$. Repeating this procedure for $M$ measurements, and still without normalizing,  we have 
\begin{align}
  P_M(H_i) \approx     \frac{1}{2^M} \exp \left( \Delta_i \sum_{n=1}^{M} x_n    \right) P_0(H_i) 
   \label{unnorm}
\end{align}
where $x_n = 1$ if the outcome of the $n^{\msi{th}}$ measurement indicates a spoof and $x_n = -1$ if it indicates a true return. Now assuming equal initial prior probabilities and normalization, the prior probabilities conditioned on the random variable $X \equiv \sum_n x_n$ are 
\begin{align}
    P_M(H_0|X) & = \frac{\exp \left( \Delta_0 X    \right) }{\exp \left( \Delta_0 X    \right)  + \exp \left( \Delta_1 X    \right) } \\
    P_M(H_1|X) & = \frac{\exp \left( \Delta_1 X    \right) }{\exp \left( \Delta_0 X    \right)  + \exp \left( \Delta_1 X    \right) } 
\end{align}
A measure of our average certainty as to which hypothesis is true is 
\begin{align}
   \langle | P_M(H_1|X)- P_M(H_0|X) | \rangle,  
   \label{uncertainty}
\end{align} 
where the average is over all possible sets of measurement outcomes $\{x_n\}$.

To evaluate this average, we need the distribution for $X$ under each hypothesis. Since $X$ is the sum of independent random variables it will be Gaussian for large enough $M$. Under the hypothesis $H_i$ this Gaussian random variable has mean and variance 
\begin{align}
    m_i & =  \sum_{n=1}^M \frac{1}{2} \left( 1 + \Delta_i   \right)  - \frac{1}{2} \left( 1 - \Delta_i  \right)  =  \sum_{n=1}^M \Delta_i = M \Delta_i \\
    V_i & =   \sum_{n=1}^M \left[ \left(1-\Delta_i\right)^2 \frac{1}{2} \left( 1 + \Delta_i   \right)  - \left(1+\Delta_i\right)^2\frac{1}{2} \left( 1 - \Delta_i   \right)   \right] \nonumber \\
    & = M \left( 1 - \Delta_i^2 \right) . 
\end{align}
The Gaussian distribution under hypothesis $H_i$ is then 
\begin{align}
    P_i(x) =  \frac{1}{\sqrt{2\pi V_i}} \exp\left[ - \frac{(x - m_i)^2 }{2 V_i} \right] 
\end{align}
The total distribution for $X$ is 
\begin{align}
    P(x) & = P_0(H_0) \frac{1}{\sqrt{2\pi V_0}} \exp\left[ - \frac{(x - m_0)^2 }{2 V_0} \right] \nonumber \\
     & +  P_0(H_1) \frac{1}{\sqrt{2\pi V_1}} \exp\left[ - \frac{(x - m_1)^2 }{2 V_1} \right] 
\end{align}
With equal initial prior probabilities, the distribution for $X$ is 
\begin{align}
    P(x) & =  \frac{1}{\sqrt{8\pi V_0}} \exp\left[ - \frac{(x - m_0)^2 }{2 V_0} \right] \nonumber \\
    & + \frac{1}{\sqrt{8\pi V_1}} \exp\left[ - \frac{(x - m_1)^2 }{2 V_1} \right] 
\end{align}
Thus our certainty measure, defined in Eq. \ref{uncertainty}, averaged over all possible measurement results is  
\begin{align}
 &\langle | P_M(H_1|X)- P_M(H_0|X) | \rangle  =  \\ 
 &\frac{1}{\sqrt{8\pi}} \int_{-\infty}^\infty \frac{e^{\Delta_1 x}-e^{\Delta_0 x} }{e^{\Delta_0 x}  + e^{\Delta_1 x} } \left(  \frac{1}{\sqrt{V_0}} \exp\left[ - \frac{(x - m_0)^2 }{2 V_0} \right]  +  \frac{1}{\sqrt{V_1}} \exp\left[ - \frac{(x - m_1)^2 }{2 V_1} \right]\right) dx  \nonumber
\end{align}
The Gaussian functions in parentheses in the integrand act as sampling functions that pick out the value of the preceding factor at $x=\Delta_0 M$ and $x=\Delta_1 M$. Then, since $\Delta_0 \approx \Delta_1$, Eq.(\ref{difference_of_priors}) follows. The notation in Sec.~\ref{Bayes} is simplified by using $P_i$ to mean $P_M(H_i|X)$. 

\backmatter


\begin{thebibliography}{10}
\expandafter\ifx\csname url\endcsname\relax
  \def\url#1{\burl{#1}}\fi
\expandafter\ifx\csname urlprefix\endcsname\relax\def\urlprefix{URL }\fi
\providecommand{\bibinfo}[2]{#2}
\providecommand{\eprint}[2][]{\url{#2}}
\providecommand{\doi}[1]{\url{https://doi.org/#1}}
\bibcommenthead

\bibitem{blakely2022quantum}
\bibinfo{author}{Blakely, J.~N.} \& \bibinfo{author}{Pethel, S.~D.}
\newblock \bibinfo{title}{Quantum limits to classically spoofing an
  electromagnetic signal}.
\newblock \emph{\bibinfo{journal}{Physical Review Research}}
  \textbf{\bibinfo{volume}{4}}, \bibinfo{pages}{023178} (\bibinfo{year}{2022}).

\bibitem{schleher1999electronic}
\bibinfo{author}{Schleher, D.~C.}
\newblock \emph{\bibinfo{title}{Electronic Warfare in the Information Age}}
  Artech House radar library (\bibinfo{publisher}{Artech House},
  \bibinfo{year}{1999}).

\bibitem{genova2018electronic}
\bibinfo{author}{Genova, J.}
\newblock \emph{\bibinfo{title}{Electronic Warfare Signal Processing}} Artech
  House electronic warfare library (\bibinfo{publisher}{Artech House},
  \bibinfo{year}{2018}).

\bibitem{strydom2012hardware}
\bibinfo{author}{Strydom, J.~J.}, \bibinfo{author}{Cilliers, J.~E.},
  \bibinfo{author}{Gouws, M.}, \bibinfo{author}{Naicker, D.} \&
  \bibinfo{author}{Olivier, K.}
\newblock \emph{\bibinfo{title}{Hardware in the loop radar environment
  simulation on wideband drfm platforms}}, \bibinfo{pages}{1--5}
  (\bibinfo{organization}{IET}, \bibinfo{year}{2012}).

\bibitem{strydom2014high}
\bibinfo{author}{Strydom, J.~J.}, \bibinfo{author}{de~Witt, J.~J.} \&
  \bibinfo{author}{Cilliers, J.~E.}
\newblock \emph{\bibinfo{title}{High range resolution x-band urban radar
  clutter model for a drfm-based hardware in the loop radar environment
  simulator}}, \bibinfo{pages}{1--6} (\bibinfo{organization}{IEEE},
  \bibinfo{year}{2014}).

\bibitem{heagney2018digital}
\bibinfo{author}{Heagney, C.~P.}
\newblock \bibinfo{title}{Digital radio frequency memory synthetic instrument
  enhancing u.s. navy automated test equipment mission}.
\newblock \emph{\bibinfo{journal}{IEEE Instrumentation \& Measurement
  Magazine}} \textbf{\bibinfo{volume}{21}}, \bibinfo{pages}{41--63}
  (\bibinfo{year}{2018}).

\bibitem{weedbrook2012gaussian}
\bibinfo{author}{Weedbrook, C.} \emph{et~al.}
\newblock \bibinfo{title}{Gaussian quantum information}.
\newblock \emph{\bibinfo{journal}{Reviews of Modern Physics}}
  \textbf{\bibinfo{volume}{84}}, \bibinfo{pages}{621} (\bibinfo{year}{2012}).

\bibitem{bennett1948spectra}
\bibinfo{author}{Bennett, W.~R.}
\newblock \bibinfo{title}{Spectra of quantized signals}.
\newblock \emph{\bibinfo{journal}{The Bell System Technical Journal}}
  \textbf{\bibinfo{volume}{27}}, \bibinfo{pages}{446--472}
  (\bibinfo{year}{1948}).

\bibitem{zamir1996lattice}
\bibinfo{author}{Zamir, R.} \& \bibinfo{author}{Feder, M.}
\newblock \bibinfo{title}{On lattice quantization noise}.
\newblock \emph{\bibinfo{journal}{IEEE Transactions on Information Theory}}
  \textbf{\bibinfo{volume}{42}}, \bibinfo{pages}{1152--1159}
  (\bibinfo{year}{1996}).

\bibitem{guctua2010quantum}
\bibinfo{author}{Gu{\c{t}}{\u{a}}, M.}, \bibinfo{author}{Bowles, P.} \&
  \bibinfo{author}{Adesso, G.}
\newblock \bibinfo{title}{Quantum-teleportation benchmarks for independent and
  identically distributed spin states and displaced thermal states}.
\newblock \emph{\bibinfo{journal}{Physical Review A}}
  \textbf{\bibinfo{volume}{82}}, \bibinfo{pages}{042310}
  (\bibinfo{year}{2010}).

\bibitem{shapiro1984phase}
\bibinfo{author}{Shapiro, J.} \& \bibinfo{author}{Wagner, S.}
\newblock \bibinfo{title}{Phase and amplitude uncertainties in heterodyne
  detection}.
\newblock \emph{\bibinfo{journal}{IEEE Journal of Quantum Electronics}}
  \textbf{\bibinfo{volume}{20}}, \bibinfo{pages}{803--813}
  (\bibinfo{year}{1984}).

\bibitem{1976quantum}
\bibinfo{author}{Helstrom, C.~W.}
\newblock \emph{\bibinfo{title}{Quantum Detection and Estimation Theory}} ISSN
  (\bibinfo{publisher}{Elsevier Science}, \bibinfo{year}{1976}).

\bibitem{helstrom1967detection}
\bibinfo{author}{Helstrom, C.~W.}
\newblock \bibinfo{title}{Detection theory and quantum mechanics}.
\newblock \emph{\bibinfo{journal}{Information and Control}}
  \textbf{\bibinfo{volume}{10}}, \bibinfo{pages}{254--291}
  (\bibinfo{year}{1967}).

\bibitem{chen2011microwave}
\bibinfo{author}{Chen, Y.-F.} \emph{et~al.}
\newblock \bibinfo{title}{Microwave photon counter based on josephson
  junctions}.
\newblock \emph{\bibinfo{journal}{Physical review letters}}
  \textbf{\bibinfo{volume}{107}}, \bibinfo{pages}{217401}
  (\bibinfo{year}{2011}).

\bibitem{pankratov2022towards}
\bibinfo{author}{Pankratov, A.} \emph{et~al.}
\newblock \bibinfo{title}{Towards a microwave single-photon counter for
  searching axions}.
\newblock \emph{\bibinfo{journal}{npj Quantum Information}}
  \textbf{\bibinfo{volume}{8}}, \bibinfo{pages}{61} (\bibinfo{year}{2022}).

\bibitem{zhuang2022ultimate}
\bibinfo{author}{Zhuang, Q.} \& \bibinfo{author}{Shapiro, J.~H.}
\newblock \bibinfo{title}{Ultimate accuracy limit of quantum pulse-compression
  ranging}.
\newblock \emph{\bibinfo{journal}{Physical review letters}}
  \textbf{\bibinfo{volume}{128}}, \bibinfo{pages}{010501}
  (\bibinfo{year}{2022}).

\bibitem{wu2022entanglement}
\bibinfo{author}{Wu, B.-H.}, \bibinfo{author}{Guha, S.} \&
  \bibinfo{author}{Zhuang, Q.}
\newblock \bibinfo{title}{Entanglement-assisted multi-aperture
  pulse-compression radar for angle resolving detection}.
\newblock \emph{\bibinfo{journal}{arXiv preprint arXiv:2207.10881}}
  (\bibinfo{year}{2022}).

\bibitem{winkler2003introduction}
\bibinfo{author}{Winkler, R.}
\newblock \emph{\bibinfo{title}{An Introduction to Bayesian Inference and
  Decision}}  (\bibinfo{publisher}{Probabilistic Publishing},
  \bibinfo{year}{2003}).

\end{thebibliography}

\end{document}